\documentclass[final]{elsart}

\usepackage{hyperref}

\usepackage{graphicx}

\usepackage{amssymb}
\usepackage{epsfig}

\begin{document}

\begin{frontmatter}



\title{Numerical extraction of de Haas--van Alphen frequencies from calculated band energies}


\author[uoft,uofb]{P.M.C.~Rourke\corauthref{corr}},
\ead{rourke@gmail.com}
\author[uoft]{S.R.~Julian}

\address[uoft]{Department of Physics, University of Toronto, Toronto, Ontario, M5S 1A7, Canada}
\address[uofb]{H. H. Wills Physics Laboratory, University of Bristol, Bristol, BS8 1TL, United Kingdom}

\begin{abstract}
A new algorithm for extracting de Haas--van Alphen frequencies and effective masses from calculated band energies is presented. The algorithm creates an interpolated $k$-space ``super cell,'' which is broken into slices perpendicular to the desired magnetic field direction. Fermi surface orbits are located within each slice, and de Haas--van Alphen frequencies and effective masses are calculated. Orbits are then matched across slices, and extremal orbits determined. This technique has been successful in locating extremal orbits not previously noticed in the complicated topology of existing UPt$_{3}$ band-structure data; these new orbits agree with experimental de Haas--van Alphen measurements on this material, and solidify the case for a fully-itinerant model of UPt$_3$.
\end{abstract}

\corauth[corr]{Correspondence to: Patrick Rourke, c/o Stephen Julian, Department of Physics, University of Toronto, Toronto, Ontario, M5S 1A7, Canada. The algorithm described herein may be downloaded as the Supercell K-space Extremal Area Finder (SKEAF) from \href{http://www.wien2k.at/reg_user/unsupported/}{\nolinkurl{http://www.wien2k.at/reg\_user/unsupported/}}.}

\begin{keyword}
dHvA \sep de Haas - van Alphen \sep Fermi surface \sep electronic structure \sep quantum oscillations \sep effective mass

\PACS 71.15.-m \sep 71.15.Dx \sep 71.18.+y \sep 71.20.-b
\end{keyword}
\end{frontmatter}

\section{Introduction}
\label{intro}

The Fermi surface of a metal is a surface of constant energy in momentum space ($k$-space) that, at absolute zero temperature, separates occupied and unoccupied electron states. Determination of the Fermi surface topology of any new metallic compound is an important step toward understanding the physics of the material, since electronic properties depend sensitively on the Fermi surface shape. A powerful experimental Fermi surface probe is measurement of the quantum oscillatory magnetization, known as the de Haas--van Alphen (dHvA) effect~\cite{shoenberg}.

At zero temperature, in the presence of a magnetic field, the free electrons of a metal undergo cyclotron motion in real space. This constrains the $k$-space electron states to lie on concentric tubes, called ``Landau tubes,'' aligned along the magnetic field direction. Only the segments of tube within the Fermi surface contain occupied states (see Fig.~\ref{fig:landautubes}). The radius of each tube depends on the strength of the magnetic field, such that as the field is increased, the tubes get larger and successively exit the Fermi surface. As each tube leaves the surface, it rapidly depopulates, causing quantum oscillations in magnetization and other properties. The quantum oscillations are periodic in $1/H$, where $H$ is the magnetic field strength, with an oscillation frequency proportional to the extremal Fermi surface cross-sectional area perpendicular to the magnetic field direction~\cite{shoenberg}:
\begin{equation}
	F = (\hbar / 2 \pi e) A
	\label{eq:dhvafreq}
\end{equation}
where $F$ is the dHvA frequency, $A$ is the extremal area, and $e$ is the elementary charge. Furthermore, as temperature is increased, the Fermi surface increasingly blurs, damping out the quantum oscillations. The damping strength depends on the effective mass of the electrons averaged around the extremal orbit being measured. Thus, dHvA measurements performed as a function of temperature and magnetic field direction directly reveal both the topology of the Fermi surface and the effective electron masses on the extremal orbits of each Fermi surface sheet~\cite{shoenberg}.

\begin{figure}[htbp]
	\centering
		\includegraphics[width=0.5\textwidth]{fig1}
	\caption{Landau tubes intersecting a spherical Fermi surface. Only the sections of tube inside the Fermi surface have occupied states. In this case, only the orbit around the equator is extremal, so only one dHvA frequency would be observed.}
	\label{fig:landautubes}
\end{figure}

\subsection{Basic concepts of Fermiology (may be skipped by readers familiar with the subject)}
\label{fermiology}

\begin{figure}[htbp]
	\centering
		\includegraphics[width=0.65\textwidth]{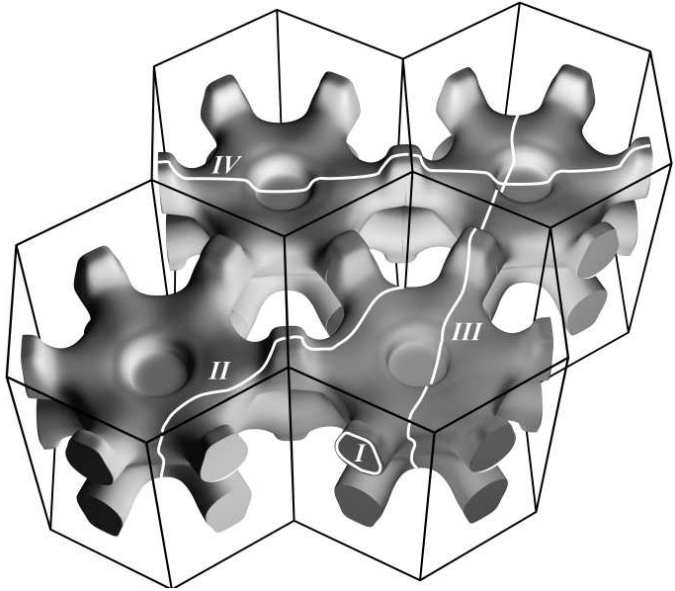}
	\caption{The band 2 Fermi surface of UPt$_{3}$, tiled in several Brillouin zones (modified from~\cite{upt3usnew}). $I$ is a simple closed orbit, easily identified as extremal by visual inspection; $II$ is a less-obvious extremal orbit that crosses Brillouin zone boundaries; $III$ is a near-open extremal orbit; and $IV$ is an open orbit.}
	\label{fig:upt3band2}
\end{figure}

For readers new to Fermi surface measurements, a few definitions are required. Since, at zero temperature, the Fermi surface separates occupied and unoccupied $k$-space states, an \textit{orbit} refers to a path along this surface that an electron may trace out under the influence of a magnetic field; in our case, we are dealing with orbits that not only lie on the constant-energy Fermi surface, but on a plane perpendicular to the applied magnetic field as well. A \textit{closed orbit} is one that forms a closed loop around some part of the Fermi surface. An \textit{extremal orbit} is a particular closed orbit whose cross-sectional area is either locally maximum or locally minimum, compared to adjacent orbits on the same Fermi surface sheet at the same magnetic field angle. Only extremal orbits are detected as quantum oscillation frequencies in a dHvA experiment; $I$, $II$ and $III$ in Fig.~\ref{fig:upt3band2} are examples of extremal orbits that occur at various angles in UPt$_{3}$.

The \textit{Brillouin zone} (or equivalently, the \textit{reciprocal unit cell}, depending on the particular volume of $k$-space enclosed) is the basic unit of momentum space, which repeats in all directions just as the real-space unit cell of a crystal does. As shown in Fig.~\ref{fig:upt3band2}, sometimes a Fermi surface links up with copies of itself in neighbouring Brillouin zones, forming one infinitely large, complicated shape. On such a surface, it is possible to have \textit{open orbits}: orbits that continue forever in one direction, never coming back on themselves to form a closed loop. $IV$ in Fig.~\ref{fig:upt3band2} is an example of an open orbit. While open orbits cannot be detected in quantum oscillation measurements, their presence can be inferred from other experiments, such as angle-resolved magnetoresistance. \textit{Near-open orbits} are closed, extremal orbits that would become open at a slightly different magnetic field angle. $III$ in Fig.~\ref{fig:upt3band2} is a near-open orbit, because a small change in magnetic field angle will transform it into an open orbit similar to $IV$. Finally, when determining the orbit type of a particular extremal orbit, an \textit{electron orbit} is one that encloses occupied states, separating them from the unoccupied states outside the orbit; conversely, a \textit{hole orbit} is one that encloses unoccupied states, and is associated with the motion of a ``hole'' through a Fermi sea of electrons rather than motion of a single electron itself.

\subsection{Comparison to band structure}
\label{intro-bandstructure}

In order to understand the physical implications of dHvA data, the measured frequencies and masses are compared to those predicted by electronic structure calculations. However, since real compounds often possess complicated Fermi surfaces, including open/near-open, nested and non-central orbits, the task of extracting predicted dHvA orbits from calculated band energies is non-trivial.

Furthermore, code for performing electronic structure calculations and generating Fermi surfaces (such as WIEN2k~\cite{wien2k}) is widely available, based on published algorithms that are used and improved by the community. In contrast, prior to this work publicly-available methods for comparing calculated Fermi surfaces to the results of quantum oscillation experiments have been scarce: while several research groups maintain private extremal area determination codes (see, for example,~\cite{prevharima,prevpietraszko,prevyelland}), their proprietary algorithms are not available to the wider community for validation, improvement or use.

One previous approach to extremal area determination that has been documented in the literature involved fitting the calculated band energies to a Fourier series of lattice-specific star functions, which was then evaluated at various points in $k$-space. An orbit centre and plane would be manually specified, then the algorithm would perform a series of ``stepping'' and ``return-to-surface'' operations as a function of rotation angle around the orbit centre to determine the cross-sectional area via Simpson's-rule integration~\cite{prevcode1969}. This approach was later expanded to included trapezoidal-rule integration for orbits which are multivalued at certain angles around the orbit centre~\cite{prevcode1975}. While such an approach works well when one knows which orbits are extremal (and the locations of the associated orbit centres), topologically-complicated Fermi surfaces can have extremal orbits that are not obvious from visual examination of the surfaces.

\section{Numerical method}
\label{nmethod}

\subsection{Overview}
\label{nmethod-overview}

Our algorithm is designed to exploit the large processing capabilities of current desktop computers in order to automatically extract extremal orbits, effective masses and density of states contributions from calculated Fermi surfaces of arbitrary topology, without requiring manual guidance or bias from the user. The code is written in the Fortran 90 language, and reads files defined in the Band-XCrysDen-Structure-File (BXSF) format. BXSF files specify band energies on a three-dimensional grid within a parallelepiped Reciprocal Unit Cell (RUC) defined by three reciprocal lattice vectors: $\vec{a}$, $\vec{b}$, and $\vec{c}$~\cite{xcrysden}. Typical input files prepared for our algorithm contain on the order of 20 000 $k$-points, whose band energies have been calculated by an electronic structure program such as WIEN2k~\cite{wien2k}.

Upon reading the input file, the following steps are performed:

\begin{enumerate}
	\item A cubic $k$-space Super Cell (SC), considerably larger than the original reciprocal unit cell and aligned with the desired magnetic field vector, is constructed (section~\ref{nmethod-supercell}). A coordinate transformation maps the super-cell $k$-point grid back into the reciprocal unit cell. Band energies at each of the super-cell grid points are determined from those provided in the reciprocal unit cell via Lagrange interpolating polynomials. The density of $k$-points in the super-cell grid is typically much greater than that of the reciprocal-unit-cell grid.
	\item The super-cell grid is divided into slices 1 $k$-point thick, perpendicular to the magnetic field vector. On each slice, the program scans through the $k$-points, locating as orbits the Fermi surface outlines (section~\ref{nmethod-fsdetect}). The cross-sectional area, effective mass and orbit type (hole or electron) are calculated for each orbit (section~\ref{nmethod-freqcalc}).
	\item Orbits are matched from slice to slice, so that each orbit is associated with a particular Fermi surface sheet (section~\ref{nmethod-omatching}).
	\item On each Fermi surface sheet, the orbits which are extremal are singled out (section~\ref{nmethod-extdet}). The orbit data for similar orbits found on separate sheets are averaged, and the results output to a file.
	\item If automatic rotation is enabled, steps 1--4 are repeated for each new magnetic field vector.
	\item The electronic density of states contribution for the band is calculated by dividing the reciprocal unit cell into a large number of equal-volume tetrahedra, determining the density of states at the Fermi level within each tetrahedron using a numerical expression, then summing the results over the entire reciprocal unit cell (section~\ref{nmethod-dos}).
\end{enumerate}

\subsection{$k$-space super cell construction}
\label{nmethod-supercell}

\begin{figure}[htbp]
	\centering
		\includegraphics[width=0.95\textwidth]{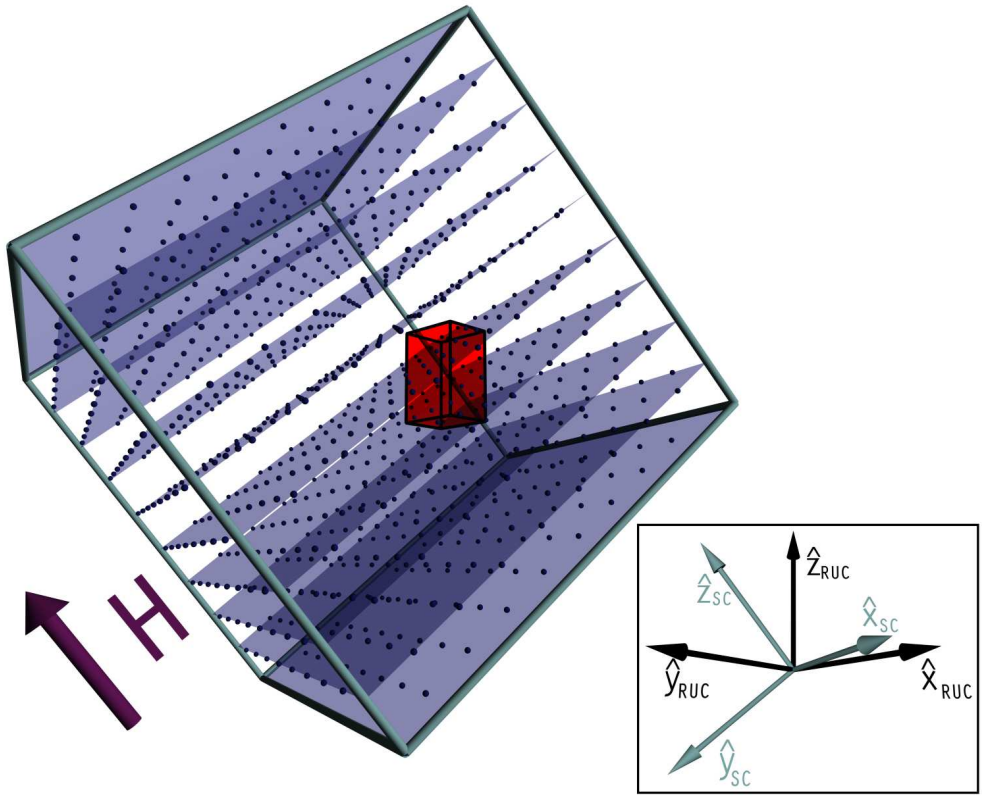}
	\caption{An example super cell, aligned with the magnetic field $H$, and drawn to-scale with the small reciprocal unit cell contained within. The super cell is cut into slices perpendicular to the magnetic field, which are populated with grid points. A typical super cell would contain 600 slices, each holding $600 \times 600$ grid points: far more than shown here. The axes for the reciprocal-unit-cell and super-cell coordinate systems are inset.}
	\label{fig:supercell}
\end{figure}

Our algorithm operates in a large cubic $k$-space Super Cell (SC), which is aligned with the magnetic field direction and contains many tiled copies of the Reciprocal Unit Cell (RUC), so the first task is to construct this cell (Fig.~\ref{fig:supercell}) and populate it with band energies.

The reciprocal lattice vectors are defined in the BXSF file relative to $k$-space axes $\hat{x}_{RUC}$, $\hat{y}_{RUC}$, $\hat{z}_{RUC}$, such that, for example:
\begin{equation}
	\vec{a} = (a_{x}, a_{y}, a_{z}) = a_{x} \hat{x}_{RUC} + a_{y} \hat{y}_{RUC} + a_{z} \hat{z}_{RUC}
	\label{eq:rucaxes}
\end{equation}
According to convention, these axes are usually chosen so that $\vec{a}$ lies along $\hat{x}_{RUC}$ (that is, $a_{y} = a_{z} = 0$); for a cubic or tetragonal reciprocal unit cell, $\vec{b}$ lies along $\hat{y}_{RUC}$ and $\vec{c}$ lies along $\hat{z}_{RUC}$. Note, however, that the super cell generation procedure can accommodate a reciprocal unit cell defined by any arbitrary reciprocal lattice vectors $\vec{a}$, $\vec{b}$ and $\vec{c}$, including those that are not orthogonal.

Since the super cell is aligned with the magnetic field ($\hat{z}_{SC} || \vec{H}$), it has a different coordinate system, and a coordinate transformation is needed to map the super-cell points back to the original reference frame:
\begin{equation}
    \left[ \begin{array}{c}
      \hat{x}_{SC} \\
      \hat{y}_{SC} \\
      \hat{z}_{SC}
      \end{array}\right]
    =
    \left[ \begin{array}{ccc}
      v^{2} u + t ~&~ -v w u ~&~ -w s \\
      -v w u ~&~ w^{2} u + t ~&~ -v s \\
      w s ~&~ v s ~&~ t
    \end{array}\right]
    \left[ \begin{array}{c}
      \hat{x}_{RUC} \\
      \hat{y}_{RUC} \\
      \hat{z}_{RUC}
    \end{array}\right]
    \label{eq:axestransform}
\end{equation}
where $s = \sin\phi$, $t = \cos\phi$, $u = 1 - \cos\phi$, $v = \sin\theta$, $w = \cos\theta$, $\phi$ is the polar angle of the magnetic field measured from $\hat{z}_{RUC}$ down toward the $\hat{x}_{RUC}$--$\hat{y}_{RUC}$ plane, and $\theta$ is the azimuthal angle of the magnetic field measured from $\hat{x}_{RUC}$ toward $\hat{y}_{RUC}$. Both coordinate systems share the same origin.

No matter what the shape of the reciprocal unit cell, the super cell is always a cube with sides defined to be 4$\times$ longer than the longest reciprocal lattice vector. This way, for any magnetic field orientation, the super cell will contain enough tiled reciprocal unit cells to be able to track orbits which cross the zone boundaries. The super cell is situated so that $1/4$ of the side length lies along $-\hat{x}_{SC}$ while $3/4$ of the side length lies along $+\hat{x}_{SC}$, and similarly for the other dimensions. The density of $k$-points in the super-cell grid is typically 100--200$\times$ greater that of the reciprocal-unit-cell grid.

The $k$-points in the super cell are transformed to the reciprocal-unit-cell reference frame via Eq.~\ref{eq:axestransform}, and translated into the volume of the original reciprocal unit cell. However, since the calculated band energies are positioned uniformly on a reciprocal-unit-cell $k$-point grid defined by the reciprocal lattice vectors, and these in turn are not necessarily at right angles to one another, a further transformation is necessary to allow the super-cell points to be compared to those read from the input file. A point $\vec{p} = p_{x} \hat{x}_{RUC} + p_{y} \hat{y}_{RUC} + p_{z} \hat{z}_{RUC}$ maps to a point $\vec{q} = q_{a} \hat{a} + q_{b} \hat{b} + q_{c} \hat{c}$ in the input array as:
\begin{equation}
    \vec{q}
    = \mathbf{M^{-1}}
    \vec{p}\textrm{, where }
    \mathbf{M} = 
    \left[ \begin{array}{ccc}
      a_{x} ~&~ b_{x} ~&~ c_{x} \\
      a_{y} ~&~ b_{y} ~&~ c_{y} \\
      a_{z} ~&~ b_{z} ~&~ c_{z}
    \end{array}\right]
    \label{eq:slantcoords}
\end{equation}

Once the super-cell $k$-points have been appropriately mapped back to the space of the input array, their band energies may be derived from those originally calculated by the electronic structure program. Since few re-mapped super-cell grid points will coincide exactly with reciprocal-unit-cell grid points, a series of third-order Lagrange interpolating polynomials~\cite{interpolation} on a $4 \times 4 \times 4$-point grid are used to determine each super-cell $k$-point band energy.

\subsection{Fermi surface orbit detection}
\label{nmethod-fsdetect}

\begin{figure}[htbp]
	\centering
		\includegraphics[width=1\textwidth]{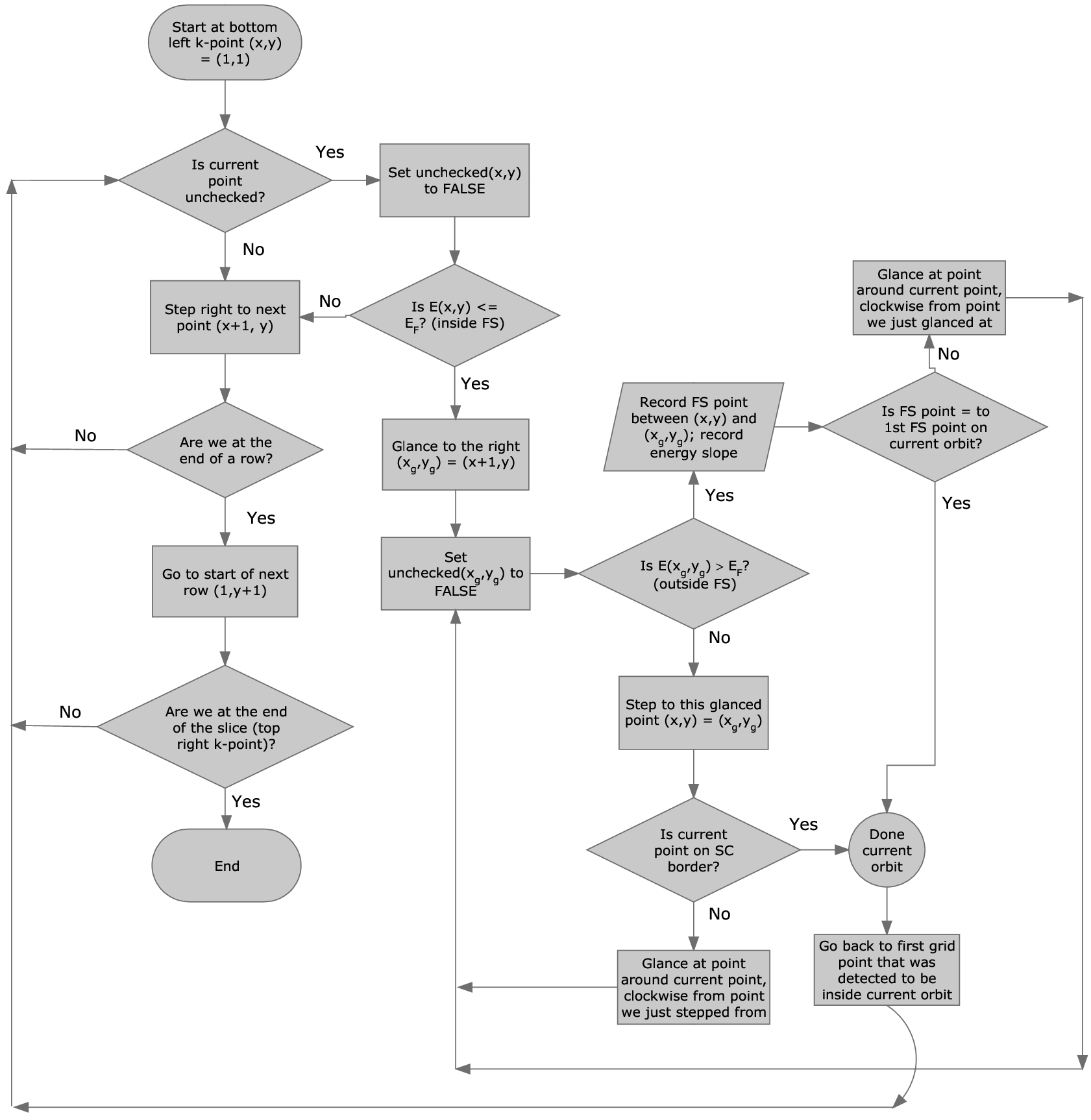}
	\caption{Flow chart showing the Fermi surface orbit detection algorithm for finding Fermi surface outlines (orbits) within a given slice. Coordinates $(x,y)$ refer to grid points in the two-dimensional array of $k$-points on the slice.}
	\label{fig:flowchart}
\end{figure}

Upon population with band energies, the super cell is cut into 1-$k$-point-thick slices perpendicular to the magnetic field direction. In each super-cell slice, the program steps through the two-dimensional $k$-point array, locating all Fermi surface orbit outlines (details of this process are shown in Fig.~\ref{fig:flowchart}). Fermi surface points and associated energy slopes are determined around each orbit. Orbits which run into the super cell boundaries are treated as open orbits and ignored (the super cell is large enough that, if the orbit is not open but merely spans a few Brillouin zones as in orbits $II$ and $III$ of Fig.~\ref{fig:upt3band2}, it will be found elsewhere).

Note that during the ``Record Fermi surface point between $(x,y)$ and $(x_{g},y_{g})$'' algorithm step (Fig.~\ref{fig:flowchart}), the Fermi surface point is not simply placed halfway between $(x,y)$ and $(x_{g},y_{g})$, but rather linearly interpolated between the two so that its position on the slice is given by:
\begin{equation} 
    \begin{array}{c}
      x_{FS,i} = x + \frac{\left| E(x,y) - E_{F} \right|}{\left| E(x,y) - E_{F} \right| ~+~ \left| E(x_{g},y_{g}) - E_{F} \right|} (x_{g} - x) \\
      y_{FS,i} = y + \frac{\left| E(x,y) - E_{F} \right|}{\left| E(x,y) - E_{F} \right| ~+~ \left| E(x_{g},y_{g}) - E_{F} \right|} (y_{g} - y) 
    \end{array}
    \label{eq:fspointpositions}
\end{equation}

The recorded energy slope is resolved into two components: $\left( \frac{d E}{d k_{||}} \right) _{i}$ parallel to the glance direction (i.e. across the Fermi surface), and $\left( \frac{d E}{d k_{\bot}} \right) _{i}$ perpendicular to the glance direction. These two orthogonal energy slope components are obtained during algorithm operation via numerical difference equations of the form:
\begin{equation} 
    \begin{array}{c}
      \left( \frac{d E}{d k_{||}} \right) _{i} = \frac{E(x_{g},y_{g}) ~-~ E(x,y)}{\Delta x (x_{g}-x) ~+~ \Delta y (y_{g}-y)} \\
      \left( \frac{d E}{d k_{\bot}} \right) _{i-1} = \frac{E(x,y) ~-~ E(x_{i-1},y_{i-1})}{\Delta x (x-x_{i-1}) ~+~ \Delta y (y-y_{i-1})} 
    \end{array}
    \label{eq:fsenergyslopes}
\end{equation}
where $\Delta x$ and $\Delta y$ are the $k$-space grid point spacings for the super cell slice $x$ and $y$ directions, respectively, and $(x_{i-1},y_{i-1})$ are the coordinates of the previous ``stepped-on'' point.

To further clarify the core Fermi surface detection algorithm presented in Fig.~\ref{fig:flowchart}, the action of this procedure on a small portion of an example slice is shown in Fig.~\ref{fig:exampleslice}. The algorithm has already been stepping around the slice by the time it arrives at the depicted portion. The operations performed, in order, are:

\begin{figure}[htbp]
	\centering
		\includegraphics[width=0.75\textwidth]{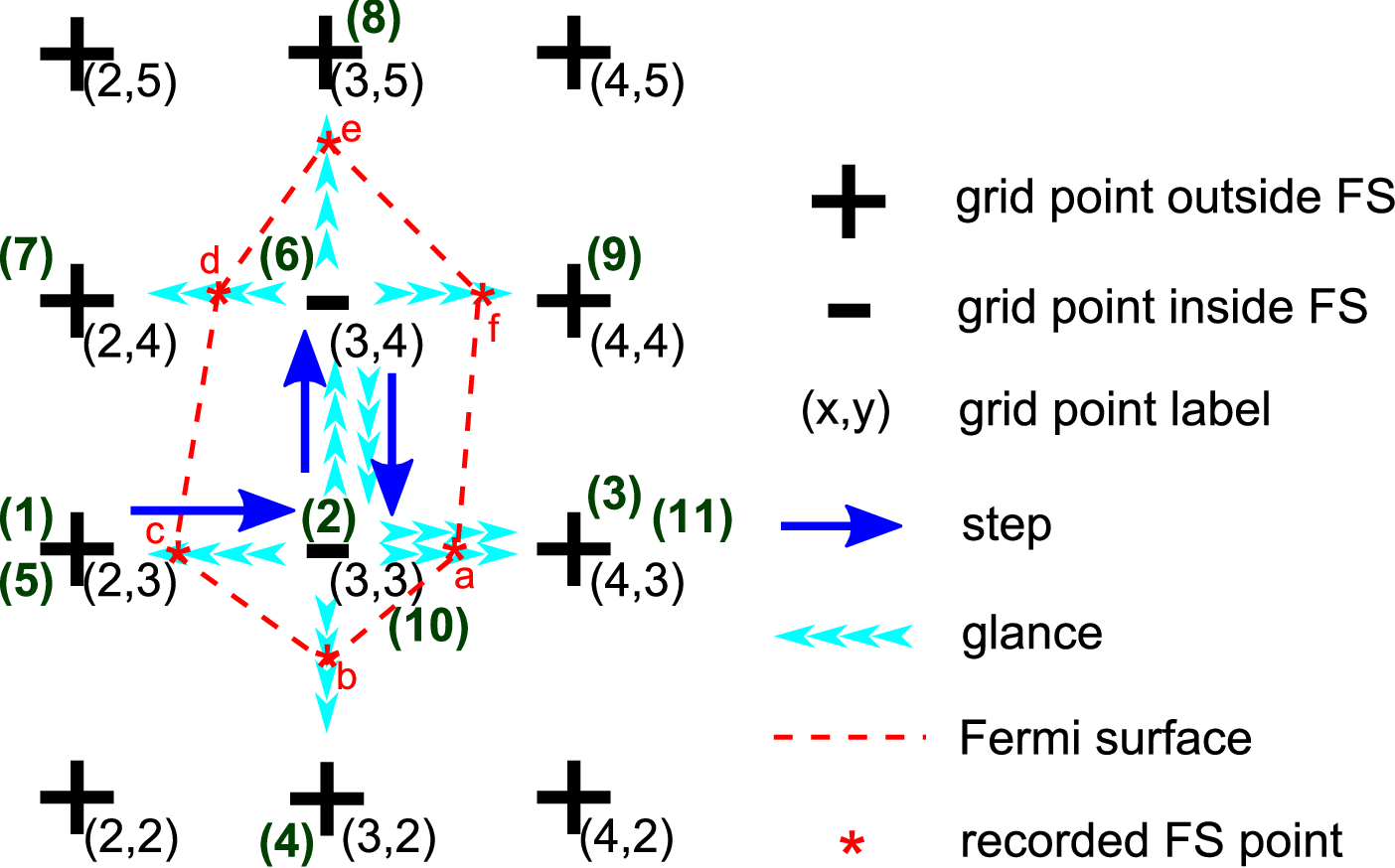}
	\caption{Small portion of an example slice. ``$+$'' grid points have energies greater than the Fermi energy; ``$-$'' grid points have energies less than the Fermi energy. Typical slices hold $600 \times 600$ grid points, so orbits usually contain many more points than the trivial example shown here.}
	\label{fig:exampleslice}
\end{figure}

\begin{enumerate}
	\item Arrive at $(2,3)$. $E(2,3)$ is not $\leq$ $E_F$.
	\item Step right to the next point, $(3,3)$. $E(3,3)$ $<$ $E_F$.
	\item Glance from $(3,3)$ to the right at $(4,3)$. $E(4,3)$ $>$ $E_F$, so Fermi surface point $a$ and the energy slope are recorded between $(3,3)$ and $(4,3)$.
	\item Glance around $(3,3)$ clockwise to $(3,2)$. $E(3,2)$ $>$ $E_F$, so Fermi surface point $b$ and the energy slope are recorded between $(3,3)$ and $(3,2)$.
	\item Glance around $(3,3)$ clockwise to $(2,3)$. $E(2,3)$ $>$ $E_F$, so Fermi surface point $c$ and the energy slope are recorded between $(3,3)$ and $(2,3)$.
	\item Glance around $(3,3)$ clockwise to $(3,4)$. $E(3,4)$ is not $>$ $E_F$, so step from $(3,3)$ to $(3,4)$.
	\item Glance around $(3,4)$ clockwise from $(3,3)$ to $(2,4)$. $E(2,4)$ $>$ $E_F$, so Fermi surface point $d$ and the energy slope are recorded between $(3,4)$ and $(2,4)$.
	\item Glance around $(3,4)$ clockwise to $(3,5)$. $E(3,5)$ $>$ $E_F$, so Fermi surface point $e$ and the energy slope are recorded between $(3,4)$ and $(3,5)$.
	\item Glance around $(3,4)$ clockwise to $(4,4)$. $E(4,4)$ $>$ $E_F$, so Fermi surface point $f$ and the energy slope are recorded between $(3,4)$ and $(4,4)$.
	\item Glance around $(3,4)$ clockwise to $(3,3)$. $E(3,3)$ is not $>$ $E_F$, so step from $(3,4)$ to $(3,3)$.
	\item Glance around $(3,3)$ clockwise from $(3,4)$ to $(4,3)$. $E(4,3)$ $>$ $E_F$, so a Fermi surface point and the energy slope are recorded between $(3,3)$ and $(4,3)$. This Fermi surface point is the same as the first one found on the orbit (Fermi surface point $a$, in step (3): this orbit is done!
\end{enumerate}

\subsection{dHvA frequency, effective mass, and orbit type calculations}
\label{nmethod-freqcalc}

For each orbit (i.e. Fermi surface outline) found in a particular slice, the dHvA frequency, effective mass and orbit type are calculated. Since the frequency is proportional to the cross-sectional Fermi surface area (Eq.~\ref{eq:dhvafreq}), it is obtained from the area of the polygon formed by the Fermi surface points (e.g. points $a$--$f$ in Fig.~\ref{fig:exampleslice}):
\begin{equation}
F = \frac{\hbar}{2 \pi e} \left| \frac{1}{2} \sum^{N - 1}_{i=1} \left( x_{FS,i} ~ y_{FS,i+1} - x_{FS,i+1} ~ y_{FS,i} \right) \right|
\label{eq:calcfreq}
\end{equation}
where the sum is over the points on the orbit, with the last point on the orbit same as the first: $ (x_{FS,N},y_{FS,N}) \equiv (x_{FS,1},y_{FS,1})$. Note that although the ``frequencies'' are calculated for every orbit, cross-sectional areas only produce experimentally-measurable dHvA frequencies when they are extremal; the extremal orbits are singled out later (section~\ref{nmethod-extdet}).

The effective mass calculation draws upon the energy slopes at the Fermi surface. As one averages around a Fermi surface outline, two different geometric cases are encountered. If the glance directions for point $i$ and point $i + 1$ are parallel (e.g. points $c$ and $d$ in Fig.~\ref{fig:exampleslice}), then:
\begin{equation}
\left| \frac{d E}{d k} \right| _{i} = \sqrt{\left( \frac{d E}{d k_{||}} \right) _{i}  ^{2} + \left( \frac{d E}{d k_{\bot}} \right) _{i}  ^{2}}
\label{eq:dedkline}
\end{equation}
If the glance directions for point $i$ and point $i + 1$ are perpendicular (e.g. points $a$ and $b$ in Fig.~\ref{fig:exampleslice}), then:
\begin{equation}
\left| \frac{d E}{d k} \right| _{i} = \sqrt{\left( \frac{d E}{d k_{||}} \right) _{i}  ^{2} + \left( \frac{d E}{d k_{||}} \right) _{i + 1}  ^{2}}
\label{eq:dedkcorner}
\end{equation}
The effective mass averaged around the orbit, given in units of the electron mass $m_{e}$, is:
\begin{equation}
m^{*} = \frac{\hbar ^2}{2 \pi m_{e}} \left( \frac{d A}{d E} \right) = \frac{\hbar ^2}{2 \pi m_{e}} \sum^{N - 1}_{i=1} \frac{\sqrt{ \left( x_{FS,i+1} - x_{FS,i} \right) ^{2} + \left( y_{FS,i+1} - y_{FS,i} \right) ^{2} }}{\left| \frac{d E}{d k} \right| _{i}}
\label{eq:effmass}
\end{equation}
Note that Eq.~\ref{eq:effmass} is only correct for extremal orbits, when the energy gradient lies within the slice.

Since the orbit-finding algorithm (section~\ref{nmethod-fsdetect}) always steps around the inside of electron orbits (e.g. Fig.~\ref{fig:exampleslice}) and the outside of hole orbits, the orbit type is determined by comparing the orbit area to the area bounded by the polygon formed by the ``stepped-on'' grid points. If the orbit area is larger than the stepped-on area, it is flagged as an electron orbit; if the reverse is true, it is flagged as a hole orbit. At this point, average coordinates (equivalent to the coordinates of the orbit centre when the orbit shape is centrosymmetric), coordinate standard deviations, maximum coordinates, and minimum coordinates for the orbit outline are also calculated.

\subsection{Slice-to-slice orbit matching}
\label{nmethod-omatching}

Once the orbit outlines have been located on all slices in the super cell, Fermi surface sheets must be reconstructed by matching orbits on adjacent slices. In order for an orbit $i$ on one slice to be matched with an orbit $j$ on the preceding slice, and therefore added to orbit $j$'s sheet, all of the following conditions must be met:

\begin{itemize}
	\item The average $x_{FS}$ and $y_{FS}$ coordinates of orbit $i$ are both within one standard deviation of the average $x_{FS}$ and $y_{FS}$ coordinates of orbit $j$.
	\item The maximum $x_{FS}$ and $y_{FS}$ coordinates of orbit $i$ are both within two standard deviations of the maximum $x_{FS}$ and $y_{FS}$ coordinates of orbit $j$.
	\item The minimum $x_{FS}$ and $y_{FS}$ coordinates of orbit $i$ are both within two standard deviations of the minimum $x_{FS}$ and $y_{FS}$ coordinates of orbit $j$.
\end{itemize}

If multiple orbits on one slice satisfy the conditions for matching with an orbit on the preceding slice, the parameter $B_{i}$ is calculated for each candidate orbit, and the orbit with the lowest $B_{i}$ value is chosen for the match:
\begin{equation}
\begin{array}{cl}
B_{i} = & \left[\textrm{avg}\left( x_{FS} \right)_{j} - \textrm{avg}\left( x_{FS} \right)_{i} \right]^{2} +  \left[\textrm{avg}\left( y_{FS} \right)_{j} - \textrm{avg}\left( y_{FS} \right)_{i} \right]^{2} \\
& + \left[\textrm{max}\left( x_{FS} \right)_{j} - \textrm{max}\left( x_{FS} \right)_{i} \right]^{2} +  \left[\textrm{max}\left( y_{FS} \right)_{j} - \textrm{max}\left( y_{FS} \right)_{i} \right]^{2} \\ 
& + \left[\textrm{min}\left( x_{FS} \right)_{j} - \textrm{min}\left( x_{FS} \right)_{i} \right]^{2} +  \left[\textrm{min}\left( y_{FS} \right)_{j} - \textrm{min}\left( y_{FS} \right)_{i} \right]^{2} 
\end{array}
\label{eq:badness}
\end{equation}

\subsection{Extremum determination}
\label{nmethod-extdet}

If an orbit has a frequency (cross-sectional area) which is greater than both of the adjacent orbits on the same sheet, or less than both of the adjacent orbits, it is taken to be extremal. Once all extremal orbits in the super cell have been selected, their average coordinates (including the $z$ coordinates taken from the positions of the super cell slices hosting the orbits) are transformed via Eqs.~\ref{eq:axestransform} and \ref{eq:slantcoords} from the super cell reference frame back to the reciprocal unit cell reference frame defined by the reciprocal lattice vectors. Extremal orbits possessing transformed average coordinates lying with a user-specified $k$-space distance from one another (default: 5\% of the reciprocal lattice vector lengths in each direction) are grouped together. Within each group, the extremal orbits are sorted by frequency, and those possessing frequencies within a user-specified fraction of the frequency of the next smallest orbit (default: 1\%) are taken to be multiple copies of the same orbit.

Since the super cell contains more than one reciprocal unit cell, it is expected that multiple copies of a given extremal orbit will be found. The copies have their transformed average coordinates, frequencies, effective masses and orbit types (electron = 1, hole = -1; useful for confirming that orbits of different type haven't been put on the same sheet) averaged, with standard deviations used to provide an indication of error bounds on these quantities.

\subsection{Density of states calculation}
\label{nmethod-dos}

In addition to extraction of dHvA frequencies and effective masses from calculated band energies, it is often useful to obtain the contribution of each band to the electronic density of states at the Fermi energy. To this end, we have implemented the well-known ``tetrahedron method'' for Brillouin zone integrations~\cite{lehmanntetra,blochltetra}. This calculation proceeds separately from the super-cell calculations, and is only performed once for a given input file (compared to the many repeated super-cell iterations for different magnetic field directions). To determine the density of states contribution, the following steps are executed:

\begin{enumerate}
	\item The band energies within the reciprocal unit cell are interpolated very finely. No super cell is used.
	\item The finely-interpolated $k$-space grid defines a series of identical parallelepiped submesh cells that fill the reciprocal unit cell. Each submesh cell has one $k$-point from the grid at each corner, and is further divided into 6 equal volume tetrahedra~\cite{blochltetra}.
	\item The $k$-space volume of the reciprocal unit cell and volume per tetrahedron, $V_{T}$, are calculated.
	\item Within each tetrahedron, the band energies at the four vertices 1--4 are sorted so that $E_1 \leq E_2 \leq E_3 \leq E_4$.
	\item The contribution of each tetrahedron to the density of states is~\cite{blochltetra}:
	\begin{equation}
DOS_{i}(E_{F}) = \left\{ \begin{array}{l}
0\textrm{,}~~~~~~~~~~~~~~~~~~~~~~~~~~\textrm{if $E_F < E_1$ or $E_F > E_4$}\\
~~~~~~~~~~~~~~~~~~~~~~~~\textrm{or $E_1 = E_2 = E_F$ or $E_3 = E_4 = E_F$}\\
~ \\
\frac{3 V_T \left( E_F - E_1 \right)^2}{\left( E_2 - E_1 \right)\left( E_3 - E_1 \right)\left( E_4 - E_1 \right)}\textrm{, if $E_1 \leq E_F \leq E_2$ and $E_1 \neq E_2$}\\
~ \\
\frac{3 V_T}{\left( E_3 - E_1 \right)\left( E_4 - E_1 \right)} \left[ 2 E_F - E_1 - E_2 + \frac{\left( E_3 - E_1 + E_4 - E_2 \right)\left( E_F - E_2 \right)^2}{\left( E_3 - E_2 \right)\left( E_4 - E_2 \right)} \right]\textrm{,}\\ 
~~~~~~~~~~~~~~~~~~~~~~~~~~~~~\textrm{if $E_2 < E_F < E_3$}\\
~ \\
\frac{3 V_T \left( E_F - E_4 \right)^2}{\left( E_4 - E_1 \right)\left( E_4 - E_2 \right)\left( E_4 - E_3 \right)}\textrm{, if $E_3 \leq E_F \leq E_4$ and $E_3 \neq E_4$}\\
\end{array} \right.
\label{eq:eldos}
\end{equation}
	\item The total band contribution to the electronic density of states at the Fermi energy is obtained by summing the tetrahedron contributions over the entire reciprocal unit cell.
\end{enumerate}

\section{Results}
\label{results}

\subsection{Test Fermi surfaces: sphere, elliptic sphere, cylinder and barrel}
\label{testfs}

In order to test our algorithm, four artificial cases with known results were constructed and used as program inputs. The input files for all test cases contain band energies specified on a $99 \times 99 \times 99$ $k$-point grid in a cubic reciprocal unit cell, with the algorithm constructing a $600 \times 600 \times 600$ $k$-point super cell at each magnetic field angle varying in one degree steps from $\phi = 0^{\circ}$ ($\vec{H} || \hat{z}_{RUC}$) to $\phi = 90^{\circ}$ ($\vec{H} || \hat{x}_{RUC}$).

\begin{figure}[htbp]
	\centering
		\includegraphics[width=\textwidth]{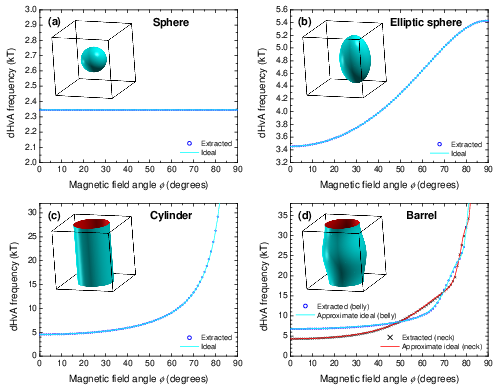}
	\caption{de Haas -- van Alphen frequency plotted as a function of polar magnetic field angle $\phi$ for the (a) sphere, (b) elliptic sphere, (c) cylinder, and (d) barrel test Fermi surfaces. For all cases but the sphere, the long axis of each Fermi surface lies along the $\phi = 0^{\circ}$ direction. Data extracted by our algorithm are shown as open symbols; ideal results for each test surface are shown as solid lines. Note that, as described in the text, the ideal curves shown in panel (d) are not exact, but rather represent an approximate analytical solution. In each panel, the inset shows the test Fermi surface within the reciprocal unit cell.}
	\label{fig:testsphcyl}
\end{figure}

First, a Fermi surface in the shape of a sphere was generated, with band energies that depend on the square of the $k$-space distance from the Fermi surface centre. This surface, shown in the inset to Fig.~\ref{fig:testsphcyl}(a), is centred at (0.5, 0.5, 0.5) in the reciprocal unit cell. An ideal sphere such as this has a single extremal orbit with the same dHvA frequency and effective mass for all magnetic field directions. In this test case, the band energies were chosen such that the dHvA frequency is $F = 2.3456$~kT and the effective mass is $m^{*} = 1.1111$~$m_e$.

Second, a Fermi surface in the shape of an elliptic sphere was generated, with band energies that depend on the square of the $k$-space distance in the $\hat{x}_{RUC}$--$\hat{y}_{RUC}$ plane from the surface's long axis and the square of the $k$-space distance (with a different amount of energy curvature) in the $\hat{z}_{RUC}$ direction from the Fermi surface centre. This surface, shown in the inset to Fig.~\ref{fig:testsphcyl}(b), is centred at (0.7, 0.6, 0.55) in the reciprocal unit cell, with the long axis aligned in the $\phi = 0$ direction. Like the sphere case above, an ideal elliptic sphere has a single extremal orbit, though now the dHvA frequency and effective mass vary with magnetic field angle as:
\begin{equation}
F \left( \phi \right) = F_{\phi = 0^{\circ}} F_{\phi = 90^{\circ}} \sqrt{ \frac{ \cot ^2 \phi + 1 }{ F^2_{\phi = 90^{\circ}} \cot ^2  \phi + F^2_{\phi = 0^{\circ}} } }
\label{eq:ellspherefreq}
\end{equation}
\begin{equation}
m^{*} \left( \phi \right) = m^{*}_{\phi = 0^{\circ}} \frac{ F \left( \phi \right) }{ F_{\phi = 0^{\circ}} }
\label{eq:ellspheremass}
\end{equation}
In this test case, the band energies were chosen such that $F_{\phi = 0^{\circ}} = 3.4567$~kT, $F_{\phi = 90^{\circ}} = 5.4321$~kT and $m^{*}_{\phi = 0^{\circ}} = 2.2222$~$m_e$.

Third, a Fermi surface in the shape of a cylinder was generated, with band energies that depend on the square of the $k$-space distance from the cylinder axis. This surface, shown in the inset to Fig.~\ref{fig:testsphcyl}(c), is centred at (0.5, 0.5, 0.5) in the reciprocal unit cell, with the cylinder axis aligned in the $\phi = 0$ direction. In an ideal cylinder, at a given magnetic field angle, the cross-sectional area is independent of $k_{z}$, with dHvA frequency and effective mass varying as:
\begin{equation}
F \left( \phi \right) = \frac{F_{\phi = 0^{\circ}}}{\cos \phi}
\label{eq:cylfreq}
\end{equation}
\begin{equation}
m^{*} \left( \phi \right) = \frac{m^{*}_{\phi = 0^{\circ}}}{\cos \phi}
\label{eq:cylmass}
\end{equation}
In this test case, the band energies were chosen such that $F_{\phi = 0^{\circ}} = 4.5678$~kT and $m^{*}_{\phi = 0^{\circ}} = 3.3333$~$m_e$. Note that the dHvA frequency asymptotes to infinity as $\phi \rightarrow 90^{\circ}$, becoming an open orbit at $\phi = 90^{\circ}$.

Finally, a Fermi surface in the shape of a barrel was generated, following the work of Yamaji~\cite{yamaji}. This surface, shown in the inset to Fig.~\ref{fig:testsphcyl}(d), is a corrugated cylinder centred at (0.5, 0.5, 0.5) in the reciprocal unit cell, with the cylinder axis aligned in the $\phi = 0$ direction. An ideal barrel will have two extremal orbits, one centred at the ``belly'' and one centred at the ``neck'' of the barrel, and in this test case the band energies were chosen such that $F_{\mathrm{belly,\phi = 0^{\circ}}} = 6.7890$~kT, $F_{\mathrm{neck,\phi = 0^{\circ}}} = 4.3210$~kT, $m^{*}_{\mathrm{belly,\phi = 0^{\circ}}} = 5.4317$~$m_e$, and $m^{*}_{\mathrm{neck,\phi = 0^{\circ}}} (\phi = 0^{\circ}) = 3.4571$~$m_e$. While we do not have an exact equation describing the dependence of the dHvA frequency and effective mass on the magnetic field angle, an approximate analytical solution (neglecting terms that are quadratic in $k_{01}$, the parameter that controls the amplitude of surface corrugation) finds that the frequency varies as~\cite{yamaji,bergemann}:
\begin{equation}
F_{\stackrel{\mathrm{belly}}{\mathrm{neck}}} \left( \phi \right) \approx \frac{\left( F_{\mathrm{belly,\phi = 0^{\circ}}} + F_{\mathrm{neck,\phi = 0^{\circ}}} \right)}{2 \cos \phi} \pm \frac{\left( F_{\mathrm{belly,\phi = 0^{\circ}}} - F_{\mathrm{neck,\phi = 0^{\circ}}} \right)}{2 \cos \phi} J_0 \left( \frac{2 \pi k_{00}}{h_{RUC}} \tan \phi \right)
\label{eq:barrelfreq}
\end{equation}
where $J_0$ is the $0^{th}$ Bessel function of the first kind, $k_{00}$ is the parameter that controls the overall size of the Fermi surface, and $h_{RUC}$ is the height of the reciprocal unit cell.

Fig.~\ref{fig:testsphcyl} shows the dHvA frequency results that our algorithm extracted from the four test cases (open symbols), plotted versus the polar magnetic field angle $\phi$; for comparison, the ideal values for each test surface, obtained using Eqs.~\ref{eq:ellspherefreq}, \ref{eq:cylfreq} and \ref{eq:barrelfreq} are also plotted (solid lines). The agreement between our extracted values and the ideal values is excellent: in the sphere, elliptic sphere and cylinder cases (for which we have exact equations governing the ideal frequency and mass angle dependence), the extracted dHvA frequencies and effective masses are within 0.05\% and 0.1\%, respectively, of their ideal counterparts over the entire angular range. In the barrel case, when $\phi = 0^{\circ}$ and we therefore know the ideal values exactly, the extracted dHvA frequencies and effective masses are likewise within 0.05\% and 0.1\%, respectively, of their ideal counterparts. In the barrel case, when $\phi > 0^{\circ}$ we no longer have an exact ideal equation and need to compare to the approximate solution of Eq.~\ref{eq:barrelfreq}; at some angles in this regime, the extracted frequencies differ from those of Eq.~\ref{eq:barrelfreq} by up to 3\%, but this is likely due to the inexact nature of the approximate ideal curve. Note that our algorithm also correctly obtains the first four Yamaji angles~\cite{yamaji} for the barrel case at $49.5^{\circ}$, $69.6^{\circ}$, $76.6^{\circ}$ and $80.1^{\circ}$---these are the magnetic field angles at which the cross-sectional area becomes independent of $k_{z}$, and therefore where the belly and neck frequencies meet.

\subsection{UPt$_3$}
\label{upt3}

Having tested our computational approach on simple Fermi surfaces, we applied it to a material of current interest: the heavy fermion superconductor UPt$_{3}$, which has a hexagonal crystal structure. The term ``heavy fermion'' is used to describe this material, because strong electron-electron correlations lead to a very large effective mass. Traditional UPt$_{3}$ band structure calculations~\cite{upt3oguchi}, in which all uranium 5$f$-electrons are itinerant rather than localized, find 5 energy bands crossing the Fermi energy~\cite{upt3normanold}, including the complicated Fermi surface sheet shown in Fig.~\ref{fig:upt3band2}. Comprehensive quantum oscillation experiments have been performed on this compound~\cite{upt3usnew}, and while the orbit shapes generally agree well with band structure, some frequencies seen experimentally had no counterparts from the calculation.

By using the previously-calculated UPt$_{3}$ band structure information~\cite{upt3normanold} as input to our program, we were able to rigorously determine the frequencies that should be seen by dHvA. Through this process, in addition to confirming the previously-known orbits, we found \textit{new extremal orbits in the old band structure calculation} that correspond to the experimentally-measured frequencies which had until now been orphaned~\cite{upt3usnew}. One such frequency, labelled $\eta$, likely corresponds to orbit $II$ in Fig.~\ref{fig:upt3band2} -- since it is difficult to see by eye that this orbit is extremal, it had not been noticed prior to our work. Alternately, the experimental $\eta$ frequency may correspond to orbit $III$ in Fig.~\ref{fig:upt3band2}, another frequency predicted by our algorithm~\cite{upt3usnew}.

\begin{figure}[htbp]
	\centering
		\includegraphics[width=1\textwidth]{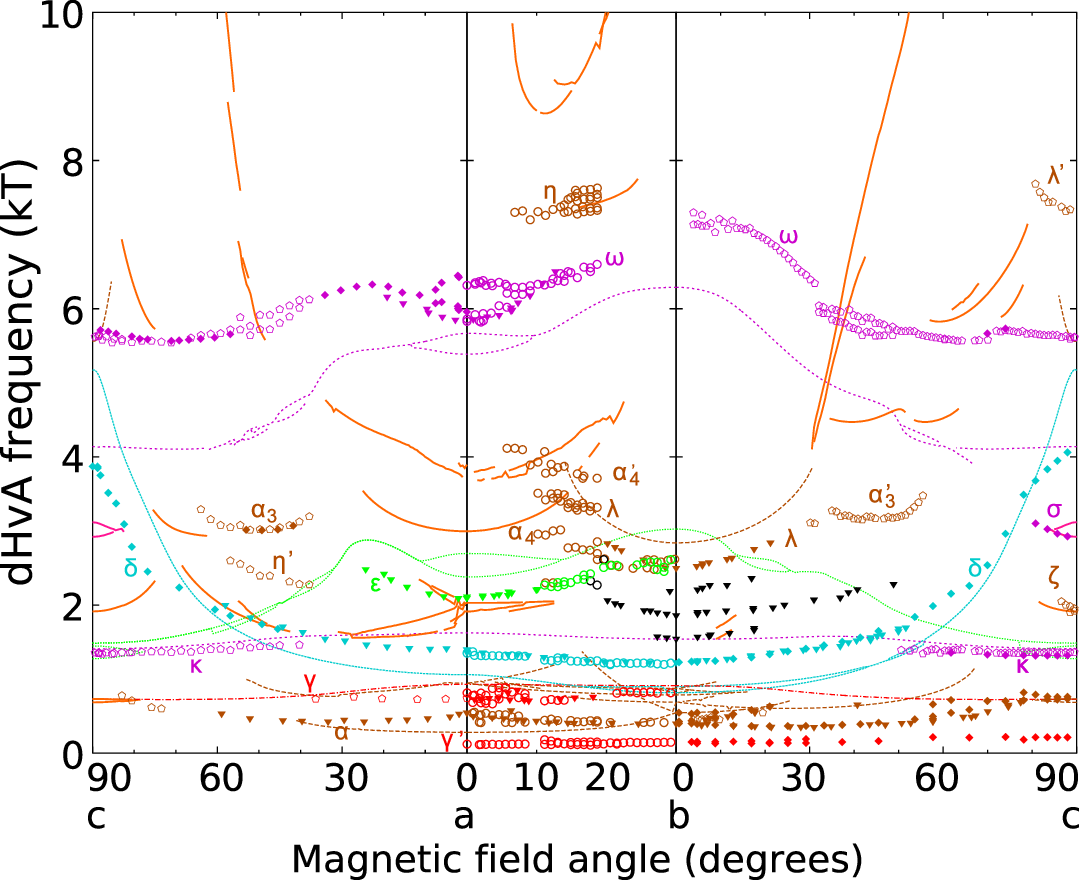}
	\caption{Measured and calculated UPt$_{3}$ de Haas -- van Alphen frequencies as a function of magnetic field angle (from~\cite{upt3usnew}). Solid and dotted lines are frequencies extracted from a fully-itinerant band structure calculation by our algorithm. Points are experimental measurements, with Greek letters indicating orbit labels. The $\eta$ orbit (orbit $II$ or $III$ in Fig.~\ref{fig:upt3band2}) is one of the new extremal orbits found in the band structure data by our algorithm.}
	\label{fig:upt3freqvsangle}
\end{figure}

Full frequency versus magnetic-field-angle results for all 5 $E_{F}$-crossing bands are shown in Fig.~\ref{fig:upt3freqvsangle}. Good agreement between measured and calculated Fermi-surface shape (manifested in the frequency variation with angle), but not size (manifested in the overall magnitude of the frequencies) is usual for heavy fermion materials, because their unusually flat bands mean that small shifts of the Fermi energy can produce large changes in the size of a Fermi surface sheet. Moreover, since quasiparticle orbits with large effective masses (as in the case with the UPt$_{3}$ Fermi surface sheet shown in Fig.~\ref{fig:upt3band2}~\cite{upt3usnew}) are difficult to detect experimentally, it is often the case that more frequencies are predicted than are measured; however, the reverse (existence of experimentally-measured frequencies without theoretically-predicted counterparts) is more serious and generally indicates a deficiency with the theoretical model. Since our calculation has located previously-missing orbits in the traditional band structure data, thus bringing it into closer agreement with the experimental results, it has helped resolve a recent controversy over the fundamental itinerant nature of the UPt$_{3}$ 5$f$ conduction electrons~\cite{upt3usnew}.

The work with UPt$_3$ provides an important further validation of our algorithm, beyond the simple test cases, because our results may be compared with predicted frequencies extracted from the same electronic structure data set using a different technique~\cite{prevcode1975,upt3normanold}. Our algorithm finds all of the UPt$_3$ extremal orbits obtained by the algorithm used by Norman \textit{et al.}, with matching frequencies and effective masses. Furthermore, the new extremal orbits discovered for the first time by our algorithm (such as orbits $II$ and $III$ in Fig.~\ref{fig:upt3band2}) are confirmed by the old algorithm when it is told where in the Brillouin zone they are located. Such agreement has also been the case with other heavy fermion compounds of current interest: in both YbRh$_2$Si$_2$ and $\beta$-YbAlB$_4$, predicted dHvA frequencies extracted from electronic structure calculations using our algorithm~\cite{ybrh2si2prl,ybrh2si2ltproc,ybalb4nevi} agree with those obtained using other theoretical techniques~\cite{ybrh2si2knebel,ybalb4tomp} and correspond to frequencies observed experimentally~\cite{ybrh2si2prl,ybrh2si2ltproc,ybrh2si2knebel,ybalb4tomp}.

\section{Conclusion}
\label{conclusion}
We have developed a new approach for extracting quantum oscillation frequency and effective mass from calculated band energies. By employing a large, heavily-interpolated $k$-space super cell and exploiting recent advances in desktop computing power, our program can robustly characterize complicated Fermi surfaces to locate all extremal orbits.

In addition to excellent performance on simple test Fermi surfaces, when applied to the complicated energy bands of UPt$_{3}$ our algorithm located new frequencies which had not been previously found on the calculated Fermi surface. These new frequencies increase the agreement between model and experiment, providing new evidence for the itinerant nature of the 5$f$ electrons in this material.

\section{Acknowledgements}
\label{acknowledgements}

The authors would like to acknowledge useful discussions with M.~R.\ Norman from Argonne National Laboratory and A.~H.\ Nevidomskyy from Rutgers University. This work is supported by the Natural Science and Engineering Research Council of Canada and the Canadian Institute for Advanced Research.



\end{document}